\documentclass[hyper]{JHEP3} 

\usepackage{axodraw,epsfig}
\usepackage{amsmath}
\usepackage{cite}

\newcommand{\be}{\begin{eqnarray}}
\newcommand{\ee}{\end{eqnarray}}
\newcommand{\bc}{\begin{center}}
\newcommand{\ec}{\end{center}}
\newcommand{\barl}{\begin{array}{rl}}
\newcommand{\barr}{\begin{array}{rr}}
\newcommand{\ball}{\begin{array}{llllll}}
\newcommand{\ea}{\end{array}}
\newcommand{\nnb}{\nonumber}
\newcommand{\bea}{\begin{eqnarray}}
\newcommand{\eea}{\end{eqnarray}}
\newcommand{\sla}{\!\!\!/}

\title{Helicity Formalism for Spin-2 Particles
       \thanks{Work supported in part by the EC 5th Framework Programme
               under contract number HPMF-CT-2002-01663.}}

\author{Tanju Gleisberg\\
        Institut f{\"u}r Theoretische Physik, TU Dresden,
        D-01062 Dresden, Germany, and\\
        Physics Department, University of Florida,
        Gainesville, FL 32611, USA\\
        E-mail: \email{tanju@theory.phy.tu-dresden.de} }
\author{Frank Krauss\\
        Theory Division, CERN, CH-1211 Geneva 23, Switzerland\\
        E-mail: \email{frank.krauss@cern.ch} }
\author{Konstantin T.~Matchev\\
        Physics Department, University of Florida,
        Gainesville, FL 32611, USA, and\\
        LEPP, Cornell University, Ithaca, NY 14853, USA\\
        E-mail: \email{matchev@phys.ufl.edu}}
\author{Andreas Sch\"alicke\\
        Institut f{\"u}r Theoretische Physik, TU Dresden,
        D-01062 Dresden, Germany\\
        E-mail: \email{dreas@theory.phy.tu-dresden.de} }
\author{Steffen Schumann\\
        Institut f{\"u}r Theoretische Physik, TU Dresden,
        D-01062 Dresden, Germany\\
        E-mail: \email{steffen@theory.phy.tu-dresden.de} }
\author{Gerhard Soff\\
        Institut f{\"u}r Theoretische Physik, TU Dresden,
        D-01062 Dresden, Germany\\
        E-mail: \email{soff@physik.tu-dresden.de}}

\received{\today}               
\accepted{\today}               

\preprint{UFIFT-HEP-03-9 \\
          CLNS-03/1829 \\
          CERN-TH/2003-135 \\ 
          hep-ph/0306182 \\ 
          August 22, 2003}

\abstract{We develop the helicity formalism for spin-2 particles
and apply it to the case of gravity in flat extra dimensions.
We then implement the large extra dimensions scenario
of Arkani-Hamed, Dimopoulos and Dvali in the program {\tt AMEGIC++},
allowing for an easy calculation of arbitrary processes
involving the emission or exchange of gravitons.
We complete the set of Feynman rules derived by Han, Lykken
and Zhang, and perform several consistency checks of our
implementation.}


\begin{document} 

\section{Introduction}

We have learned from string theory that extra spatial dimensions
are the price for unifying the Standard Model with gravity.
Yet we are practically ignorant about how and why these
extra dimensions are hidden from our world. Will they appear
at the electroweak scale? Are they bosonic or fermionic? 
We hope that the answers to these puzzling questions
are awaiting us in the next round of collider experiments at
the Tevatron, LHC, and, possibly, at the next linear collider.

If spatial extra dimensions exist and are accessible to at least
some of the Standard Model fields, they have to be very small,
$\le{\cal O} ({\rm TeV}^{-1})$. However, it is conceivable that
all Standard Model particles are confined to a brane, 
and only gravity (which is rather poorly tested at short
distances) is allowed to permeate the bulk.
This is the classic scenario of large extra dimensions
\cite{Arkani-Hamed:1998rs,Antoniadis:1998ig},
from now on referred to as ADD.
In ADD the experimental limits on the size of the
extra dimensions are much weaker. For recent
compilations of various experimental probes of large extra
dimensions, see e.g. \cite{Landsberg:2001ma,Hewett:2002hv,Hewett:cx}.

Such scenarios offer an alternative solution to the gauge hierarchy problem
and new avenues for unifying the Standard Model interactions with
gravity. They also predict
a variety of novel signals, which are already testable at high-energy colliders,
see \cite{Cheung:2003ah} and references therein.
Generally speaking, there are two classes of signatures:
real graviton production 
\cite{Mirabelli:1998rt,Giudice:1998ck,Han:1998sg,Cheung:1999zw,Balazs:1999ge,Han:1999ne}
and virtual graviton exchange \cite{Giudice:1998ck,Han:1998sg,Hewett:1998sn}. 
Gravitons have enormous lifetime and
decay off the brane, hence real graviton production
results in generic missing energy signatures.
Virtual graviton exchange may generate numerous higher 
dimension operators, contributing to the production
of fermion~\cite{Giudice:1998ck,Han:1998sg,Hewett:1998sn,fermion},
gauge boson~\cite{GB,Cheung:1999ja,Lee:1999ft} or 
Higgs boson~\cite{Rizzo:1999qv,He:1999ay} pairs, as well as more
complicated final states with more than two particles
\cite{Dvergsnes:2002nc,Deshpande:2003sy}.

The phenomenology of ADD extra dimensions has been studied 
rather extensively, and numerous experimental
bounds from LEP and the Tevatron on the scale of extra
dimensions $M_s$ have been set, yet none of the
existing general purpose Monte Carlo event generators has
fully and consistently implemented the ADD model.
For example, {\tt PYTHIA} \cite{Sjostrand:2001yu}
now allows only for the production of narrow
graviton resonances as in Randall--Sundrum-type models
\cite{Randall:1999ee,Randall:1999vf}. 
Virtual graviton exchange is not implemented.
For Run I analyses \cite{Acosta:2002eq,Abazov:2003gp,monojetsCDF}
of large extra dimensions, the Tevatron collaborations
have been using an unofficial version of {\tt PYTHIA},
where ADD graviton production has been implemented
as an external process \cite{LM}. 
The situation with {\tt HERWIG} \cite{Corcella:2002jc}
is very similar -- narrow graviton resonances
are included \cite{Allanach:2000nr,Allanach:2002gn},
but not much else. 
{\tt ISAJET} \cite{Baer:1999sp} has recently added the
process of real ADD graviton production in association
with a jet or a photon \cite{Vacavant:sd} and the
implementation is analogous to \cite{LM}.
{\tt PANDORA} \cite{Peskin:1999hh}
has implemented virtual graviton exchange for
$e^+e^-\to f\bar{f}, \gamma \gamma, ZZ, W^+W^-$
with the contact interaction of \cite{Hewett:1998sn}
as well as $e^+e^-\to \gamma G$ as in \cite{Mirabelli:1998rt}.
Finally, programs such as {\tt Madgraph} \cite{Maltoni:2002qb}
or {\tt COMPHEP} \cite{Pukhov:1999gg}, although
highly automated, do not contain spin-2 particles and thus
cannot be applied for graviton studies in their present form.
Given all this, the pressing need for an automated 
event generator, capable of studying ADD phenomenology
in its full generality, becomes quite evident.

The ADD Feynman rules have appeared in
\cite{Giudice:1998ck,Han:1998sg}
\footnote{See, however, section \ref{sec:vertex_rules}.}, 
so it may seem that
implementing the ADD model in one of the above-mentioned
event generators is rather straightforward. However,
apart from a few simple
cases of two-body final states, the calculations 
involve a large number of diagrams, whose evaluation
through the standard textbook methods of squaring the
amplitudes, employing the completeness relations
and taking the traces, becomes prohibitively
difficult. This problem is 
considerably alleviated when the helicity method is used,
which avoids lengthy matrix 
multiplications stemming from the notorious chains of 
Dirac matrices. First ideas pointing into this 
direction can be found in \cite{Jacob:at,Bjorken:1966kh}; 
later on, different approaches were formulated in
\cite{DeCausmaecker:1981by,DeCausmaecker:1981bg,Berends:1981uq,Gastmans:xh,Caffo:ds,Passarino:1983bg,Passarino:1983zs,Berends:gf,Kleiss:1985yh,Kleiss:1986qc,Hagiwara:1985yu} and further 
refined in \cite{Ballestrero:1992dv,Ballestrero:1994ti,Ballestrero:1994jn}.
Thanks to their easy implementation and automatization
(see for instance \cite{Tanaka:1989gu,Murayama:1992gi,Stelzer:1994ta})
helicity methods have become a cornerstone for modern 
programs dealing with the evaluation of cross sections for 
multi-particle final states. Examples for such programs are 
the multi-purpose matrix element generators {\tt GRACE} 
\cite{Yuasa:1999rg}, {\tt HELAC} \cite{Kanaki:2000ey}, 
{\tt O'MEGA} \cite{Ohl:2000hq}, {\tt AMEGIC++} 
\cite{Krauss:2001iv}, {\tt ALPGEN} \cite{Mangano:2002ea}, 
and {\tt Madgraph} \cite{Maltoni:2002qb}; furthermore 
there is a large number of more specialized codes such
as those described in 
\cite{Berends:1994xn,Accomando:2002sz,Dittmaier:2002ap}.

The main purpose of this paper is to develop
the helicity formalism for spin-2 particles and to implement 
the ADD scenario in {\tt AMEGIC++} 
(A Matrix Element Generator In C++).
Previous studies of ADD extra dimensions using helicity amplitudes
can be found in~\cite{Han:1999ne,Peskin:1999hh}.

The outline of the paper is as follows. In Section \ref{sec2} we review the 
basics of the helicity formalism and discuss some ideas
(\ref{SecS1Prop} and \ref{SecS2Prop}) that allow a general extension 
to the case of spin-2 particles (gravitons). 
In Section~\ref{sec:implementation} we
specify to the case of ADD and highlight some features of 
the specific implementation in {\tt AMEGIC++}.
For the main part we follow the conventions of \cite{Han:1998sg}.
We describe both external and virtual graviton processes and
list a couple of Feynman rules that had not been published before. 
Section~\ref{sec:tests} compares this implementation 
to some existing results in the literature,
while Section~\ref{sec:conclusions} contains our conclusions.
Several novel studies of ADD phenomenology at current and
future colliders will be presented in a separate publication
\cite{future}. 

\section{The Helicity Method}
\label{sec2}

The basic idea behind the helicity 
method is to replace Dirac matrices by suitable spinor 
products, and to use the fact that expressions like
\bea\nnb
\bar u(\lambda_1,\, p_1) u(\lambda_2,\,p_2)
\eea
are simple scalar functions of the helicities $\lambda$
and the momenta $p$ of the spinors. 

\subsection{Spinor basis}
The spinor basis used in {\tt AMEGIC++} follows closely the
definitions of \cite{Ballestrero:1994ti,Ballestrero:1994jn},
where spinors are defined in such a way that they satisfy the 
Dirac equations
\bea\label{Dirac}
p\sla u(p) = +m\cdot u(p)\;,\;\;
p\sla v(p) = -m\cdot v(p)\,,
\eea
with 
\bea\nnb
p\sla^2 = p^2 = m^2\,,
\eea
where $m^2$ is not constrained to be positive. For space-like
$p$, i.e.\ $p^2 < 0$, $m$ is purely imaginary, and the resulting
phase degeneracy can be lifted by considering eigenstates of
$\gamma_5s\sla$. With the polarization vector $s$ satisfying
\bea
s\cdot p = 0\;,\;\; s^2 = -1\,,
\eea
this translates into fixing the degeneracy by the conditions
\bea\label{Spins}
\gamma_5s\sla u(p,\pm) = \pm u(p,\pm)\;,\;\;
\gamma_5s\sla v(p,\pm) = \pm v(p,\pm)\,.
\eea
Spinors can now be constructed explicitly along the lines
of \cite{Kleiss:1985yh}. Defining two orthogonal
auxiliary vectors $k_{0,1}$ with
\bea
k_0^2 = 0\;,\;\; k_1^2 = -1\;,\;\; k_0\cdot k_1 = 0\;,
\eea
massless spinors $w$ can be defined through
\bea
w(k_0,\lambda)\bar w(k_0,\lambda) = 
\frac{1+\lambda\gamma_5}{2}k\sla_0\,,
\eea
and the relative phase between spinors with different 
helicity is given by
\bea
w(k_0,\lambda) = \lambda k\sla_1 w(k_0,-\lambda)\,.
\eea
Spinors for an arbitrary four-momentum $p$ with $p^2 = m^2$ 
can be obtained from the massless ones with the help of
\bea
u(p,\lambda) = 
\frac{p\sla+m}{\sqrt{2p\cdot k_0}} w(k_0,-\lambda)\;,\;\;
v(p,\lambda) = 
\frac{p\sla-m}{\sqrt{2p\cdot k_0}} w(k_0,-\lambda)\,,
\eea
both satisfying Eqs.~(\ref{Dirac}) and (\ref{Spins})
\footnote{In \cite{Andreev:2001se} spinors for massive fermions 
          have been proposed which are directly related to 
          physical polarization, i.e spin, states.}.
In this approach, conjugate spinors are defined to
satisfy relations similar to those of Eq.~(\ref{Spins}),
\bea\label{Spins1}
\bar u(p,\pm) \gamma_5s\sla = \pm \bar u(p,\pm)\;,\;\;
\bar v(p,\pm) \gamma_5s\sla = \pm \bar v(p,\pm)\,,
\eea
and their relation to the massless spinors is given by
\bea
\bar u(p,\lambda) = \bar w(k_0,-\lambda) 
\frac{p\sla+m}{\sqrt{2p\cdot k_0}}\;,\;\;
\bar v(p,\lambda) = \bar w(k_0,-\lambda)
\frac{p\sla-m}{\sqrt{2p\cdot k_0}}\,.
\eea
This guarantees the orthogonality of the spinors and after
normalizing
\bea
\frac{\bar u(p,\lambda) u(p,\lambda)}{2m} = 1\;,\;\;
\frac{\bar v(p,\lambda) v(p,\lambda)}{2m} = 1\;,
\eea
the following completeness relation holds true:
\bea
1 = \sum_\lambda\frac{u(p,\lambda)\bar u(p,\lambda)-
                      v(p,\lambda)\bar v(p,\lambda)}{2m}\,.
\eea

\subsection{Fermion lines}
In particular, the numerators of fermion propagators 
experience the following manipulation
\bea
(p\sla+m) = 
\frac12\sum\limits_{\lambda}
\left[\left(1+\frac{m}{\sqrt{p^2}}\right) 
           u(p,\lambda)\bar u(p,\lambda)  +
      \left(1-\frac{m}{\sqrt{p^2}}\right) 
           v(p,\lambda)\,\bar v(p,\lambda) 
\right]\,,
\eea
which is valid also for space-like $p$, i.e.\ for $p^2<0$.
For a more detailed derivation, see \cite{Ballestrero:1994jn}. 
This replacement allows us to ``cut'' diagrams into pieces with no internal 
fermion lines.

\subsection{External gauge bosons}

For spin-1 bosons there are two different strategies to deal with 
the polarization vectors:
\begin{enumerate}
\item either replace them by spinor products, or
\item define explicit polarization 4-vectors.   
\end{enumerate}
Both options are available in {\tt AMEGIC++}.

\subsubsection{Replacing polarization vectors by spinor products}
\label{Pol2Spinor}
In this section, the replacement of polarization vectors of
external massless or massive spin-1 bosons by spinor products is 
briefly summarized.
\begin{itemize}
\item Massless spin-1 bosons on their mass-shell have two physical 
      degrees of freedom; hence their polarization vectors satisfy
      \bea\label{Pol0}
      \ball
      \epsilon_\mu(p,\lambda) \, p^\mu &=& 0 \,, &
      \epsilon_\mu(p,\lambda) \, \epsilon^\mu(p,\lambda) &=& 0  \,, \\
      \epsilon_\mu(p,-\lambda) &=& \epsilon_\mu^{*}(p,\lambda)  \,, &
      \epsilon_\mu(p,\lambda) \, \epsilon^\mu(p,-\lambda) &=& -1\,,
      \ea
      \eea
      where $p$ and $\lambda$ are the four-momentum and
      the polarization label of the boson, respectively.
      In the axial gauge the polarization sum reads
      \bea\label{PolSum0}
      \sum_{\lambda=\pm} 
      \epsilon_\mu(p,\lambda)\epsilon_\nu^{*}(p,\lambda) =
      -\eta_{\mu\nu} + \frac{p_\mu q_\nu+p_\nu q_\mu}{p\cdot q}\,,
      \eea 
      where the auxiliary vector $q \nparallel p$ and $\eta_{\mu\nu}$
      denotes the flat metric tensor.
      Then, it can be shown that the replacement of the polarization 
      vectors through
      \bea
      \epsilon_\mu(p,\,\lambda) = 
      \frac{1}{\sqrt{4p\cdot q}}\,
      \bar u(q,\lambda)\gamma_\mu u(p,\lambda)
      \eea
      satisfies all relations of Eqs.~(\ref{Pol0}) and (\ref{PolSum0})
      and is thus a good choice. 
\item For massive vector bosons one chooses a similar strategy to construct 
      a spinor expression replacing the polarization vectors.
      However, massive spin-1 bosons on their mass-shell have an
      additional degree of freedom, the two transverse polarizations
      are supplemented with a longitudinal one with $\lambda = 0$. 
      This state obeys
      \bea\label{PolM}
      \epsilon_\mu(p,0) \, p^\mu = 0 \;,\;\;&
      \epsilon_\mu(p,0) \, \epsilon^\mu(p,\lambda) = 0 \;,\;\;
      \epsilon_\mu(p,0) \, \epsilon^\mu(p,0) = -1 \,,
      \eea      
      and the corresponding spin sum becomes
      \bea\label{PolSumM}
      \sum_{\lambda=\pm,0} 
      \epsilon_\mu(p,\lambda)\epsilon_\nu^{*}(p,\lambda) =
      -\eta_{\mu\nu} + \frac{p_\mu p_\nu}{m^2}\,,
      \eea
      where $p$ and $m$ are the four-momentum and the mass
      of the boson. In general, there is no simple replacement
      for the polarization vectors as for the massless case.
      However, for unpolarized cross sections, the only relevant
      requirement is that Eq.~(\ref{PolSumM}) be reproduced. 
      This can be simply achieved by introducing vectors and
      kinematics related to a pseudo-decay of the boson into two
      massless spinors with momenta $r_{1,2}$ satisfying
      \bea
      r_1^2 = r_2^2 = 0 \;,\;\;
      r_1^\mu + r_2^\mu = p^\mu\,.
      \eea
      Obviously the quantity 
      \bea
      a_\mu = \bar u(r_2,-) \gamma_\mu u(r_1,-)
      \eea
      projects out any scalar off-shell polarization. This quantity
      can be used to replace the polarization vector, because
      the integral over the solid angle $\Omega$ of $r_1$
      available for the decay leads, up to a constant normalization
      factor, to the spin sum of Eq.~(\ref{PolSumM}),
      \bea
      \frac{3}{8\pi m^2}\int d\Omega\, a_\mu a_\nu^{*} = 
       - \eta_{\mu\nu} + \frac{p_\mu p_\nu}{m^2}\,. 
      \eea
      Therefore, replacing
      \bea
      \epsilon_\mu \to a_\mu \;,\;\;
      \sum_{\lambda=\pm,0} 
      \epsilon_\mu\epsilon_\nu^{*} \to
      \frac{3}{8\pi m^2}\int d\Omega\, a_\mu a_\nu^{*} 
      \eea
      leads to the desired result. A first discussion of these
      ideas can be found in \cite{Passarino:1983zs}.
\end{itemize}

\subsubsection{Polarization basis}\label{PolBasis}
It often turns out to be more convenient to keep the polarizations explicitly.
A possible choice for an on-shell vector boson with momentum
\bea
p_{\mu} = \left(p_{0},\left|\vec{p}\right|\sin\theta\cos\varphi,
\left|\vec{p}\right|\sin\theta\sin\varphi,\left|\vec{p}\right|\cos\theta\right)
\nonumber
\eea
is
\bea\label{exPol}
\ball
\epsilon_{\mu}^\pm(p) &=& \frac{1}{\sqrt{2}}\left(0,
\cos\theta\cos\varphi \mp i\sin\varphi,\cos\theta\sin\varphi \pm i\cos\varphi,
-\sin\theta \right) \\
\epsilon_{\mu}^0(p) &=& \frac{1}{\sqrt{p^{2}}}\left(\left|\vec{p}\right|,
p_{0} \frac{\vec{p}}{\left|\vec{p}\right|}\right).
\ea
\eea
For the special choice of $q=\left(p_{0},-\vec{p}\right)$ 
this satisfies Eq.~(\ref{Pol0}) and Eq.~(\ref{PolSum0}) for a massless boson;
for massive bosons, also Eq.~(\ref{PolM}) and Eq.~(\ref{PolSumM}) are fulfilled.

Note that from now on we will label the polarization mode with an upper index.
To avoid confusion,
Lorentz indices are always denoted by $\mu,\nu,\rho$ or $\sigma$, while 
$\lambda$ and $\kappa$ are used to indicate a vector polarization mode.

\subsection{Gauge boson propagators}
\label{SecS1Prop}
Similar to the external vector boson treatment, there are two strategies to 
deal with gauge boson propagators\footnote{
This refers to the soon-to-be released version 2.0 of {\tt AMEGIC++}.}:
\begin{itemize}
\item Structures in Feynman amplitudes with internal boson lines are 
      calculated as one function (composite building block), depending 
      just on the spinors at its edges. 
      (External vector bosons are replaced as in \ref{Pol2Spinor}.)
\item The numerator of vector boson propagators can be replaced by products 
      of polarization vectors similar to the completeness relations in 
      Eq.~(\ref{PolSum0}) and Eq.~(\ref{PolSumM}). This allows us to ``cut''
      internal boson lines. 
\end{itemize}

The replacement of the numerator for a massive spin-1 boson in unitary 
gauge reads 
\begin{eqnarray}\label{s1Prop}
\sum_{\lambda = t_1,t_2,l,s} \epsilon_{\mu}^\lambda(p) 
\epsilon_{\nu}^\lambda(p)
=-\eta_{\mu\nu}+\frac{p_{\mu}p_{\nu}}{m^{2}},
\end{eqnarray}
where $p_{\mu} = \left(p_{0},\left|\vec{p}\right|\sin\theta\cos\varphi,
\left|\vec{p}\right|\sin\theta\sin\varphi,\left|\vec{p}\right|\cos\theta\right)$ 
can now be off-shell. The corresponding polarization vectors are
\begin{eqnarray}
\epsilon_{\mu}^{t_1}(p) &=& \left(0,\cos\theta\cos\varphi,
\cos\theta\sin\varphi,-\sin\theta \right) \,, \nonumber \\
\epsilon_{\mu}^{t_2}(p) &=& \left(0,-\sin\varphi,\cos\varphi,0 \right) \,, \nonumber \\
\epsilon_{\mu}^l(p) &=& \frac{1}{\sqrt{p^{2}}}\left(\left|\vec{p}\right|,
p_{0} \frac{\vec{p}}{\left|\vec{p}\right|}\right) \,, \nonumber \\
\epsilon_{\mu}^s(p) &=& \sqrt{\frac{p^{2}-m^{2}}{p^{2}m^{2}}}p_{\mu}\,,
\end{eqnarray}
with the physical boson mass $m$.
To simplify the implementation
we choose a basis with linear polarization vectors 
$\epsilon_{\mu}^{t_1}$ and $\epsilon_{\mu}^{t_2}$ for the transverse modes.
Together with the longitudinal mode $\epsilon_{\mu}^{l}$ these vectors 
yield Eq.~(\ref{PolSumM}) for particles on their mass-shell.
They are normalized and orthogonal.  
The scalar polarization $\epsilon_{\mu}^{s}$ reflects the 
``off-shellness'' of the propagating particle.
The orthogonality to all other polarization vectors is obvious, 
since it is proportional to $p_{\mu}$.
Note that in Eq.~(\ref{s1Prop}) there is no complex conjugation in the
polarization product. This avoids an extra sign discussion for
space-like $p$, where $\epsilon_{\mu}^{l}$ and $\epsilon_{\mu}^{s}$ may 
become imaginary.
The above replacement is also valid for vector bosons with a width, if $m^{2}$ 
is substituted by $m^{2}+im\Gamma$.

A simple inspection of Eq.~(\ref{s1Prop}) shows that a similar relation for 
massless vector bosons can be obtained by taking the limit
$m\rightarrow\infty$. Then, only $\epsilon_{\mu}^{s}$ is affected  
and turns out to be
\begin{eqnarray}
\epsilon_{\mu}^{s_0}(p) &=& \sqrt{-\frac{1}{p^{2}}}\ p_{\mu}.
\end{eqnarray}
By contracting the corresponding Lorentz index in each of the 
neighbouring vertices with a polarization vector
and summing over the degrees of freedom, the Lorentz structure of the 
propagator is absorbed.

\subsection{Spin-2 bosons}
\label{SecS2Prop}
In this section we will show how the above ideas can be extended
to deal with spin-2 particles. We follow the conventions and Feynman rules
derived in \cite{Han:1998sg}. With some minimal changes, this can be applied to
other models involving quantum gravity as well
\cite{Randall:1999ee,Randall:1999vf}.

The propagator for a KK-graviton with mass $m$ is
\begin{eqnarray}
i\triangle_{\{\mu\nu\},\{\rho\sigma\}}(p) =
\frac{i B_{\mu\nu,\rho\sigma}(p)}
{p^{2}-m^{2}+i\varepsilon}, 
\end{eqnarray}
where
\begin{eqnarray}
B_{\mu\nu,\rho\sigma}(p) & = & \frac{1}{2} 
\left(\eta_{\mu\rho}-\frac{p_{\mu}p_{\rho}}{m^{2}}\right)
\left(\eta_{\nu\sigma}-\frac{p_{\nu}p_{\sigma}}{m^{2}}\right)
+\frac{1}{2}
\left(\eta_{\mu\sigma}-\frac{p_{\mu}p_{\sigma}}{m^{2}}\right)
\left(\eta_{\nu\rho}-\frac{p_{\nu}p_{\rho}}{m^{2}}\right)
\nonumber \\
& & -\frac{1}{3}
\left(\eta_{\mu\nu}-\frac{p_{\mu}p_{\nu}}{m^{2}}\right)
\left(\eta_{\rho\sigma}-\frac{p_{\rho}p_{\sigma}}{m^{2}}\right).
\label{Bmunurhosigma}
\end{eqnarray}
Employing Eq.~(\ref{s1Prop}) for each term in parentheses, we obtain
\begin{eqnarray}\label{s2PropSym}
B_{\mu\nu,\rho\sigma} &=& 
\sum_{\lambda,\kappa = t_1,t_2,l,s}\left( \frac{1}{2}
\epsilon_{\mu}^\lambda \epsilon_{\rho}^\lambda 
\epsilon_{\nu}^\kappa \epsilon_{\sigma}^\kappa
+ \frac{1}{2}
\epsilon_{\mu}^\lambda \epsilon_{\sigma}^\lambda
\epsilon_{\nu}^\kappa \epsilon_{\rho}^\kappa - \frac{1}{3}
\epsilon_{\mu}^\lambda \epsilon_{\nu}^\lambda 
\epsilon_{\rho}^\kappa \epsilon_{\sigma}^\kappa\right).
\end{eqnarray}
Since the graviton couples to the energy--momentum tensor, 
all resulting vertices are symmetric in $\mu\leftrightarrow\nu$. 
Using this symmetry the above equation can be simplified to
\begin{eqnarray}\label{s2Prop}
B_{\mu\nu,\rho\sigma} = 
\sum_{\lambda,\kappa = t_1,t_2,l,s}\left(
\epsilon_{\mu}^\lambda \epsilon_{\rho}^\lambda 
\epsilon_{\nu}^\kappa \epsilon_{\sigma}^\kappa
- \frac{1}{3}
\epsilon_{\mu}^\lambda \epsilon_{\nu}^\lambda 
\epsilon_{\rho}^\kappa \epsilon_{\sigma}^\kappa\right).
\end{eqnarray}
The polarization vectors $\epsilon^{\lambda}$ have been 
defined in the previous section. 
To absorb the Lorentz structure of a spin-2 propagator into the vertices,
their Lorentz indices now have to be contracted by two polarization vectors 
and the double sum in Eq.~(\ref{s2Prop}) has to be performed. 

External spin-2 particles can be described by polarization tensors.
A possible choice is given in \cite{Han:1998sg} \footnote{Notice that
our normalization convention differs from the last version of
\cite{Han:1998sg} by a factor of $\sqrt{2}$. We chose to do this
in order for the polarization sums to be normalized to 1, as in
Eq.~(A.3) in the published version of \cite{Han:1998sg}.}:
\begin{equation}
\epsilon_{\mu\nu}^{k}=\left\{\epsilon_{\mu}^{+}\epsilon_{\nu}^{+},
\frac{1}{\sqrt{2}}\left(\epsilon_{\mu}^{+}\epsilon_{\nu}^{0}+
\epsilon_{\mu}^{0}\epsilon_{\nu}^{+}\right),
\frac{1}{\sqrt{6}}\left(\epsilon_{\mu}^{+}\epsilon_{\nu}^{-}+
\epsilon_{\mu}^{-}\epsilon_{\nu}^{+}-2\epsilon_{\mu}^{0}\epsilon_{\nu}^{0}\right),
\frac{1}{\sqrt{2}}\left(\epsilon_{\mu}^{-}\epsilon_{\nu}^{0}+
\epsilon_{\mu}^{0}\epsilon_{\nu}^{-}\right),
\epsilon_{\mu}^{-}\epsilon_{\nu}^{-}\right\}.
\end{equation}
Again applying the $\mu\leftrightarrow\nu$ symmetry in the vertices,
this simplifies to
\begin{eqnarray}
\epsilon_{\mu\nu}^{k}=\left\{\epsilon_{\mu}^{+}\epsilon_{\nu}^{+},
\sqrt{2}\,\epsilon_{\mu}^{+}\epsilon_{\nu}^{0}\,,
\sqrt{\frac{2}{3}}\left(\epsilon_{\mu}^{+}\epsilon_{\nu}^{-}-
\epsilon_{\mu}^{0}\epsilon_{\nu}^{0}\right),
\sqrt{2}\,\epsilon_{\mu}^{-}\epsilon_{\nu}^{0}\,,
\epsilon_{\mu}^{-}\epsilon_{\nu}^{-}\right\}.
\end{eqnarray}
Here, $\epsilon_{\mu}^{\pm,0}$ are the usual polarization vectors for massive
vector bosons, which were defined in Section \ref{PolBasis}.
All polarization tensors are normalized and orthogonal, and they 
fulfill the completeness relation
\begin{eqnarray}\label{s2completeness}
\sum_{k = 1...5} \epsilon_{\mu\nu}^{k} \epsilon_{\rho\sigma}^{k}
= B_{\mu\nu,\rho\sigma} \;.
\end{eqnarray}

\subsection{Construction of building blocks}
\label{buildingblocks}

Utilizing the manipulations discussed above, all possible Feynman 
amplitudes can be built up from a small set of basic building blocks.
Denoting all spinors by $u$, defining chirality projectors $P_{L,R}$
through
\bea
P_{L,R} = \frac{1\mp\gamma_5}{2}
\eea
and defining
\bea
\eta_i = \sqrt{2 p_i\cdot k_0}\;,\;\;
\mu_i  = \pm \frac{m_i}{\eta_i},
\eea
where $+/-$ refers to particle/antiparticle,
the basic building blocks are as follows:
\begin{enumerate}
\item
Basic functions are direct products of massless spinors of the form
\bea
S(\lambda;p_1,p_2) \equiv \bar u(p_1,\lambda) u(p_2,-\lambda)\,.
\eea
Expressed through the auxiliary vectors $k_{0,1}$, they read
\bea
S(\pm;p_1,p_2) &=& 
   2\frac{\pm(p_1\cdot k_0)(p_2\cdot k_1)\mp
          (p_1\cdot k_1)(p_2\cdot k_0)-
          i\epsilon_{\mu\nu\rho\sigma}
          p_1^\mu p_2^\nu k_0^\rho k_1^\sigma}{\eta_1\eta_2}\,,\nnb\\
S(\pm;p_1,p_2) &=& -S(\pm;p_2,p_1)\,.
\eea

\item
The simplest structure is the direct product of two (possibly massive) 
spinors of the form
\bea
Y(p_1,\lambda_1;p_2,\lambda_2;{c_R,c_L}) \equiv
\bar u(p_1,\lambda_1)\left[c_R P_R + c_L P_L\right]u(p_2,\lambda_2)\,,
\eea
which emerges for instance through couplings of scalar bosons
to a fermion current. The results for this function are summarized 
in Table \ref{Yfuncs}.
\begin{table}
\begin{center}
\begin{tabular}{|c|c||c|c|} \hline
$\lambda_1\lambda_2$ & $Y(p_1,\lambda_1;p_2,\lambda_2;c_L,c_R)$ &
$\lambda_1\lambda_2$ & $Y(p_1,\lambda_1;p_2,\lambda_2;c_L,c_R)$\\ \hline
$++$ & $c_R\mu_1\eta_2 + c_L\mu_2\eta_1$ & $+-$ & $c_LS(+;p_1,p_2)$\\ \hline
\end{tabular}
\end{center}
\caption{\label{Yfuncs} $Y$-functions for different helicity combinations.
         Missing combinations can be obtained through the 
         replacements $+ \leftrightarrow -$ and $L \leftrightarrow R$.}
\end{table}

\item 
The next basic structure is the product of two spinors with
a Dirac matrix in between. This emerges for instance when external
gauge bosons couple to fermion lines with the polarization vector of the
gauge bosons expressed explicitly or from terms $\sim p^\mu p^\nu$ of 
massive gauge boson propagators. This building block is called
$X$-function and reads 
\bea
X(p_1,\lambda_1;p_2;p_3,\lambda_3;c_L,c_R) = 
\bar u(p_1,\lambda_1) p\sla_2[c_LP_L+c_RP_R]u(p_3,\lambda_3)\,.
\eea
Only the restriction $k_0p_2\neq 0$ has to be applied by a suitable
choice for $k_0$. The results for this function are summarized in
Table \ref{Xfuncs}. 
\begin{table}
\begin{center}
\begin{tabular}{|c|c|}\hline
$\lambda_1\lambda_3$ & $X(p_1,\lambda_1;p_2;p_3,\lambda_3;c_L,c_R)$\\ \hline
$++$ & $ \mu_1\mu_3\eta_2^2c_L+\mu_2^2\eta_1\eta_3c_R 
+c_RS(+;p_1,p_2)S(-;p_2,p_3)$ \\
$+-$ & $c_L\mu_1\eta_2S(+;p_2,p_3)+c_R\mu_3\eta_2S(+;p_1,p_2)$\\ \hline
\end{tabular}
\end{center}
\caption{\label{Xfuncs} $X$-functions for different helicity combinations.
         Missing combinations can be obtained through the 
         replacements $+ \leftrightarrow -$ and $L \leftrightarrow R$.}
\end{table}

\item
The last structure to be discussed is
\bea
\lefteqn{Z(p_1,\lambda_1;p_2,\lambda_2; p_3,\lambda_3;p_4,\lambda_4;
           c_L^{12},c_R^{12};c_L^{34},c_R^{34}) =} \nnb\\
&&
\bar u(p_1,\lambda_1)\gamma^\mu [c_L^{12}P_L+c_R^{12}P_R]u(p_2,\lambda_2)
\bar u(p_3,\lambda_3)\gamma_\mu [c_L^{34}P_L+c_R^{34}P_R]u(p_4,\lambda_4)\,,
\eea
which stems, for example, from terms $\sim \eta^{\mu\nu}$ in gauge boson
propagators between fermion lines. Results for this function can be found in
Table~\ref{Zfuncs}.
\begin{table}
\begin{center}
\begin{tabular}{|c|c|}\hline
$\lambda_1\lambda_2\lambda_3\lambda_4$ & $Z(p_1,\lambda_1;p_2,\lambda_2;
p_3,\lambda_3;p_4,\lambda_4; c_L^{12},c_R^{12};c_L^{34},c_R^{34})$\\ \hline
$++++$ & $2\left[S(+;p_3,p_1)S(-;p_2,p_4)c_R^{12}c_R^{34}+ \mu_1\mu_2\eta_3
\eta_4c_L^{12}c_R^{34}+ \mu_3\mu_4\eta_1\eta_2c_R^{12}c_L^{34}\right]$\\
$+++-$ & $2\eta_2c_R^{12}\left[S(+;p_1,p_4)\mu_3c_L^{34} +S(+;p_1,p_3)
\mu_4c_R^{34}\right]$\\
$++-+$ & $2\eta_1c_R^{12}\left[S(-;p_3,p_2)\mu_4c_L^{34} +S(-;p_4,p_2)
\mu_3c_R^{34}\right]$\\
$++--$ & $2\left[S(+;p_4,p_1)S(-;p_2,p_3)c_R^{12}c_L^{34}+ \mu_1\mu_2\eta_3
\eta_4c_L^{12}c_L^{34}+ \mu_3\mu_4\eta_1\eta_2c_R^{12}c_R^{34}\right]$\\
$+-++$ & $2\eta_4c_R^{34}\left[S(+;p_1,p_3)\mu_2c_R^{12} +S(+;p_2,p_3)
\mu_1c_L^{12}\right]$\\
$+-+-$ & $0$\\
$+--+$ & $-2\left[ \mu_1\mu_4\eta_2\eta_3c_L^{12}c_L^{34} +\mu_2\mu_3
\eta_1\eta_4c_R^{12}c_R^{34} -\mu_1\mu_3\eta_2\eta_4c_L^{12}c_R^{34}
-\mu_2\mu_4\eta_1\eta_3c_R^{12}c_L^{34}\right]$\\
$+---$ & $2\eta_3c_L^{34}\left[S(+;p_4,p_2)\mu_1c_L^{12} +S(+;p_1,p_4)
\mu_2c_R^{12}\right]$\\ \hline
\end{tabular}%
\end{center}
\caption{\label{Zfuncs} $Z$-functions for different helicity combinations.
         Missing combinations can be obtained through the 
         replacements $+ \leftrightarrow -$ and $L \leftrightarrow R$.}
\end{table}
\end{enumerate}

\section{Implementation of the ADD Model}
\label{sec:implementation}

In our implementation of the ADD model we have applied the 
formalism developed in \cite{Han:1998sg}. We assume
$n$ flat extra dimensions of common size $R$, compactified 
on an $n$-dimensional torus. While the size $R$ is unambiguous,
it has become customary in the literature to quote bounds 
on the size of the extra dimensions in terms of an energy
scale $M$, for which there are several popular conventions
\cite{Giudice:1998ck,Han:1998sg,Hewett:1998sn}.
We choose to work with the scale $M_S$ 
as defined in the revised version of \cite{Han:1998sg}:
\begin{equation}
R^{n}=\frac{(4\pi)^{n/2}\Gamma(n/2)}{2M_S^{n+2}G_N},
\label{msdef}
\end{equation}
where $G_N=1/(8\pi \bar{M}_P^2)$ is the 4-dimensional Newton's constant,
with $\bar{M}_P=2.4\times10^{18}$ GeV the reduced Planck scale.
The other commonly used scale, $M_D$, introduced in \cite{Giudice:1998ck}
is related to $M_S$ as
\begin{equation}
M_S = 2\sqrt{\pi}\ \left[\Gamma(n/2)\right]^{1/(n+2)} M_D.
\end{equation}
Virtual graviton exchange processes are sometimes parametrized 
in terms of higher dimension operators whose coefficients are
suppressed by a scale $\Lambda_T$ \cite{Giudice:1998ck} or
$M_S^{JH}$ \cite{Hewett:1998sn}.
The relation to $M_S$ can be obtained, comparing
(\ref{leadingterm}) with (\ref{DGRW}) and (\ref{DJH}).

\subsection{Summation of the KK states}
While the coupling of matter particles to
a single graviton is suppressed by the inverse of the 
4-dimensional Planck mass $\bar{M}_P$, 
the amplitude of any given process involving gravitons is
enhanced by the multiplicity of graviton states.
After summing over the tower of the graviton mass states,
the effective suppression is only of order $M_S$.

\subsubsection{Virtual KK particles}
\label{sec:virtual}

The amplitude for any process involving a virtual graviton exchange
contains the sum 
\begin{eqnarray}
D(s) &=& \sum_{\vec{n}}\frac{i}{s-m^{2}_{\vec{n}}+i\varepsilon}\ , 
\label{Ds}
\end{eqnarray}
which is most easily calculated by replacing the sum over states of discrete
mass 
\begin{equation}
m^{2}_{\vec{n}}=\frac{4\pi^{2}\vec{n}^{2}}{R^{2}}
\label{mn}
\end{equation}
by an integral over a continuous mass distribution\footnote{Note 
that in principle there is an additional
dependence on $m^2_{\vec{n}}$ due to the mass-dependent terms in
Eq.~(\ref{Bmunurhosigma}). 
However, these mass-dependent terms vanish completely if the particles 
interacting with the graviton are all on-shell. Even if this is not 
the case, the error resulting from neglecting the additional 
$m_{\vec{n}}$ dependence in Eq.~(\ref{Ds}) is rather small -- 
our numerical tests showed deviations mostly of the order of the 
numerical precision ($\approx 10^{-14}$) and in some extreme cases 
up to $10^{-5}$. An effective mass could be chosen in such a
way that the contribution of the $\frac{1}{m_{\vec{n}}^{2}}$ term 
is correct.}~\cite{Han:1998sg}.
The result for time-like propagators reads
\begin{eqnarray}
D(s) &=& \frac{s^{n/2-1}}{2M_{s}^{n+2}G_{N}}\left[\pi+2iI(M_{s}/\sqrt{s})\right]\,, 
\label{timelike}
\end{eqnarray}
and for space-like propagators it is given by
\begin{eqnarray}
D_{E}(t) &=& \frac{|t|^{n/2-1}}{2M_{s}^{n+2}G_{N}}(-2i)I_{E}(M_{s}/\sqrt{|t|}),
\label{spacelike}
\end{eqnarray}
with
\begin{eqnarray}
I(x) & = & \left\{ \begin{array}{ll}
 \displaystyle - \sum_{k=1}^{n/2-1}\frac{1}{2k}x^{2k} 
       - \frac{1}{2}\log\left(x^{2}-1\right)
                & \quad n=\rm{even}  \\[5mm]
 \displaystyle - \sum_{k=1}^{(n-1)/2}\frac{1}{2k-1}x^{2k-1}
       + \frac{1}{2}\log\left(\frac{x+1}{x-1}\right)
                & \quad n=\rm{odd} \,,
                   \end{array} \right.
\end{eqnarray}
\begin{eqnarray}
I_{E}(x) & = & \left\{ \begin{array}{ll}
 \displaystyle  (-1)^{n/2+1}\left[\sum_{k=1}^{n/2-1}\frac{(-1)^{k}}{2k}x^{2k}
 + \frac{1}{2}\log\left(x^{2}+1\right)\right]
              & \quad n=\rm{even} \label{I} \\[5mm]
 \displaystyle  (-1)^{(n-1)/2}\left[ \sum_{k=1}^{(n-1)/2}\frac{(-1)^{k}}{2k-1}x^{2k-1}
  +  \tan^{-1}(x)\right]
              & \quad  n=\rm{odd} \,.
   \end{array} \right.
\label{IE}
\end{eqnarray}
In Eqs.~(\ref{timelike}) and (\ref{spacelike}) above,
we have used Eq.~(\ref{msdef}).
In both cases, the leading order terms for \(M_{s}^{2}\gg s\) are:
\begin{eqnarray}
D_{E}(|t|\rightarrow s) \approx D(s) &\approx& 
  \left\{ \begin{array}{ll} 
  \displaystyle \frac{-i}{2M_{s}^{4}G_{N}}\log\left(\frac{M_{s}^{2}}{s}\right) 
         & \quad n=2  \\[5mm]
  \displaystyle \frac{-i}{2M_{s}^{4}G_{N}}\frac{2}{n-2}
         & \quad n>2 \,.
   \end{array}\right.
\label{leadingterm}
\end{eqnarray}

By default, we have implemented in the program the complete
expressions (\ref{timelike}--\ref{IE}). However, for
the sake of comparison with previous work on the subject,
we have also implemented the leading order approximation
(\ref{leadingterm}), as well as the two other
scale conventions mentioned earlier.
In the GRW convention \cite{Giudice:1998ck},
the sum (\ref{Ds}) amounts to
\begin{eqnarray}\label{DGRW}
D = \frac{-i}{2\Lambda_{T}^{4}G_{N}},
\end{eqnarray}
while in the Hewett convention \cite{Hewett:1998sn}, it reads
\begin{eqnarray}\label{DJH}
D = \frac{-i\lambda}{\pi (M^{JH}_{s})^{4}G_{N}},
\end{eqnarray}
where the unspecified sign $\lambda$ of the amplitude 
reflects the arbitrariness in the regularization procedure
employed in obtaining (\ref{timelike}-\ref{IE}).

\subsubsection{External KK particles}
\label{sec:external}

For real graviton production processes, we must 
sum over all possible final state gravitons.
A single mass state is labeled by the $n$-dimensional vector 
$\vec n = (n_1,n_2,...,n_n)$, where the $n_i$ are integer values.
The corresponding mass $m_{\vec n}$ is given by Eq.~(\ref{mn}).
The total number of accessible graviton mass states 
in the KK tower is
\bea
\frac{1}{n}\frac{\hat{m}^{n}}{M_s^{n+2}G_N},
\label{GTM}
\eea
with $\hat{m}$ the maximal kinematically allowed 
graviton mass. The scale $M_S$ is defined by Eq.~(\ref{msdef}).
Since this number is 
enormous, it is impractical to introduce
a separate particle label in the program 
for each graviton of definite mass. 
Instead, it is easier to treat the graviton tower
as a single particle of variable mass, and then add an
extra summation over the graviton mass \cite{LM,Vacavant:sd}.
In this respect, real graviton production is unlike
any of the conventional processes implemented in the
general purpose event generators, where
the particle masses are held fixed.

The summation over the graviton mass can be easily 
carried out by the Monte Carlo phase-space 
integrator used in \cite{Krauss:2001iv}. 
Any given scattering event produces a single
graviton whose mass $m_{\vec n}$ is picked at random
from a uniform distribution of $\vec n$.
Thus, to get the total cross section,
the weight of every phase space point has to be 
re-weighted by the factor (\ref{GTM}).

\subsection{Vertex Feynman Rules}
\label{sec:vertex_rules}

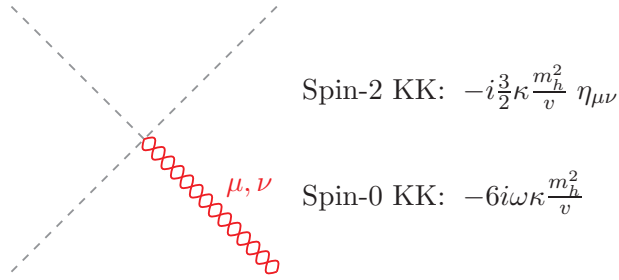
\begin{figure}[ht]
\begin{center}
{
\unitlength=1.0 pt
\SetScale{1.0}
\SetWidth{0.7}      
\normalsize    
{} \qquad\allowbreak
\begin{picture}(300,100)(0,0)
\SetColor{Gray}
\DashLine(0,100)(50,50){3}
\DashLine(100,100)(0,0){3}
\SetColor{Red}
\Photon(50,50)(100,0){3}{7}
\Photon(50,50)(100,0){-3}{7}
\Text(90,32)[c]{\Red{$\mu,\nu$}}
\Text(110,70)[l]{\Black{Spin-2 KK: $\; -i\frac{3}{2}\kappa\frac{m_{h}^{2}}{v}\;
\eta_{\mu\nu}$}}
\Text(110,30)[l]{\Black{Spin-0 KK: $\; -6i\omega\kappa\frac{m_{h}^{2}}{v}$}}
\end{picture}
}
\caption{\label{3Higgs} The triple-Higgs--KK vertex.}
\end{center}
\end{figure}

\begin{figure}[ht]
\begin{center}
{
\unitlength=1.0 pt
\SetScale{1.0}
\SetWidth{0.7}      
\normalsize    
{} \qquad\allowbreak
\begin{picture}(300,100)(0,0)
\SetColor{Gray}
\DashLine(0,100)(100,0){3}
\DashLine(100,100)(0,0){3}
\SetColor{Red}
\Photon(50,50)(50,0){3}{5}
\Photon(50,50)(50,0){-3}{5}
\Text(55,0)[l]{\Red{$\mu,\nu$}}
\Text(110,70)[l]{\Black{Spin-2 KK: $\; -i\frac{3}{2}\kappa\frac{m_{h}^{2}}{v^2}\;
\eta_{\mu\nu}$}}
\Text(110,30)[l]{\Black{Spin-0 KK: $\; -6i\omega\kappa\frac{m_{h}^{2}}{v^2}$}}
\end{picture}
}
\caption{\label{4Higgs} The four-Higgs--KK vertex.}
\end{center}
\end{figure}
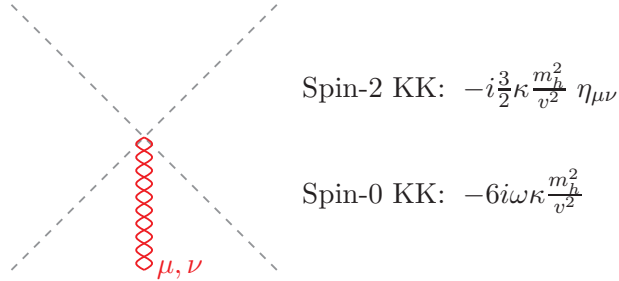

We have implemented all three- and some of the four-point
vertices from \cite{Han:1998sg} for both the spin-2 gravitons
and the spin-0 dilaton. We also included a triple-Higgs--KK
vertex, which together with a four-Higgs--KK vertex, was not 
previously listed in the literature, to the best of our knowledge. 
The Feynman rules for these two additional vertices 
are shown in Figs.~\ref{3Higgs} and \ref{4Higgs}.\\

Appendix \ref{Vertex} contains a list of all implemented vertex Feynman rules, 
translated  into the functions defined as building blocks in 
Section~\ref{buildingblocks}.

Note that we define the coupling constants $\kappa$ and 
$\omega$ somewhat differently from \cite{Han:1998sg}. This
should not become an issue, since $\kappa$ and $\omega$ are
auxiliary variables, and the ADD model is defined 
entirely through $M_S$ and $n$.
In the revised version of \cite{Han:1998sg}, the spin-2 propagator
and the normalization of the polarization tensors have been changed by 
a factor of 2, which has been absorbed into $\kappa$.
We have also rescaled $\omega$ by an extra factor of $\sqrt{n-1}$, 
which typically arises from the multiplicity of the spin-0 KK particles
\begin{eqnarray}
\kappa = \sqrt{32\pi G_{N}},\;\;\;\;
\omega = \sqrt{\frac{n-1}{3(n+2)}}.
\end{eqnarray}
With these definitions the vertex Feynman rules can be simply
read off from either the published or revised version of
\cite{Han:1998sg} and we do not repeat them here.

\section{Tests}
\label{sec:tests}

As a test of our implementation we performed a number
of checks, comparing {\tt AMEGIC++} results with
independent analytical calculations in the literature:
\begin{itemize}
\item In order to test 2-fermion production processes mediated by
      virtual graviton exchange, we calculated the
      angular distributions for 
      $e^{+}e^{-}\rightarrow \mu^{+}\mu^{-}$,
      $e^{+}e^{-}\rightarrow b\bar{b}$ and
      $e^{+}e^{-}\rightarrow c\bar{c}$. We found each one in
      good agreement with the original work of \cite{Hewett:1998sn}.
\FIGURE[ht]{\epsfig{file=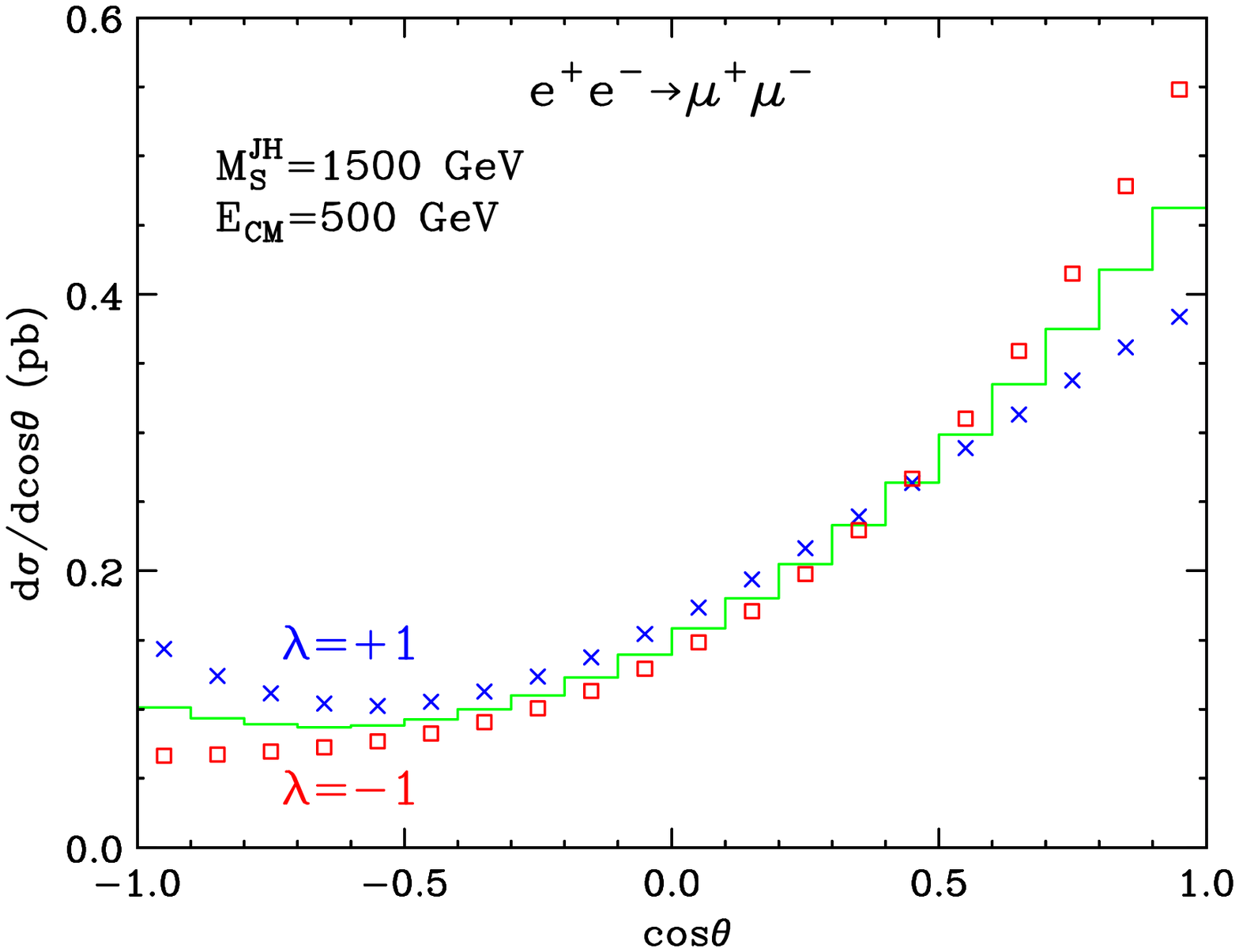,width=10cm} 
\caption{The {\tt AMEGIC++} result for
$e^{+}e^{-}\rightarrow \mu^{+}\mu^{-}$, 
for centre-of-mass energy $E_{CM}=500$ GeV,
$M_S^{JH}=1500$ GeV, and $\lambda=+1$ (crosses) 
or $\lambda=-1$ (squares). The solid histogram is the
pure Standard Model result.}
\label{muons}}
      Fig. \ref{muons} shows the {\tt AMEGIC++} result for
      $e^{+}e^{-}\rightarrow \mu^{+}\mu^{-}$, 
      for a centre-of-mass energy $E_{CM}=500$ GeV.
      We ran {\tt AMEGIC++} in the effective operator mode
      corresponding to the Hewett convention and chose
      $M_S^{JH}=1500$ GeV. Fig. \ref{muons} displays our results
      for both $\lambda=+1$ (crosses) and $\lambda=-1$ (squares),
      as well as the pure Standard Model result (solid histogram).
      We see perfect agreement with Fig.~1 of the archived version of
      \cite{Hewett:1998sn}. We also ran {\tt AMEGIC++} in its default
      mode, choosing the appropriate $M_S$ according to Eq.~(\ref{DJH})
      and recovered the curve corresponding to $\lambda=+1$, thus
      confirming the result of \cite{Landsberg:2001ma}.
\item The unpolarized angular distribution for
      $e^{+}e^{-}\rightarrow W^{+}W^{-}$ was calculated
      and found to be consistent with the 
      results of \cite{Lee:1999ft}.
\item All of our results on $\gamma\gamma\rightarrow hh$ were found
      to be in perfect agreement with those of Ref.~\cite{He:1999ay}.
\item The total cross sections for
      $\gamma\gamma\rightarrow\gamma\gamma$ and
      $e^{+}e^{-}\rightarrow\gamma\gamma$ were
      calculated and compared with the results of \cite{Cheung:1999ja}.
      We find agreement, except for the case $n=2$ in 
      $\gamma\gamma\rightarrow\gamma\gamma$. 
      We believe that the reason for the deviation is that
      \cite{Cheung:1999ja} used the approximation
      $|t|\sim s$ for $t$-channel diagrams, as in Eq.~(\ref{leadingterm}), 
      while the more general $t$-channel result is given by 
      Eq.~(\ref{spacelike}).
\item We tested real graviton production through
      $e^{+}e^{-}\rightarrow\gamma G$.
\FIGURE[ht]{\epsfig{file=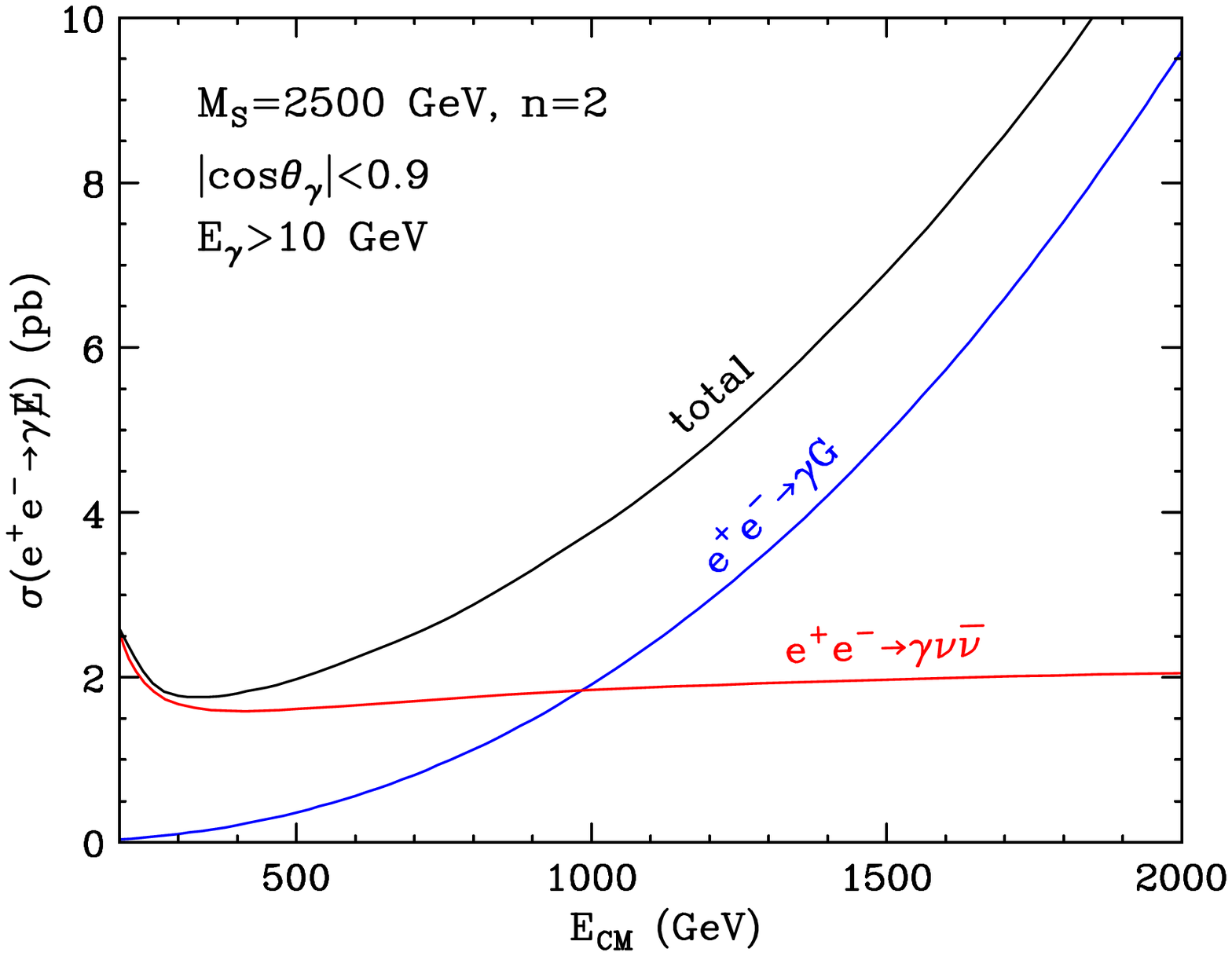,width=10cm} 
\caption{The total cross-section for $e^{+}e^{-}\rightarrow\gamma G$
as a function of the centre-of-mass energy $E_{CM}$,
for $M_S=2500$ GeV, $n=2$ and 
photon energy and angular acceptance cuts of $E_\gamma>10$ GeV and
$|\cos\theta_\gamma|<0.9$, respectively. Compare with
Fig.~3 of \cite{Cheung:1999zw}.}
\label{photon}}
      Our results for the total
      cross section as a function of the centre-of-mass energy 
      $E_{CM}$ are shown in Fig.~\ref{photon} for $M_S=2500$ GeV, $n=2$, and 
      for photon energy and angular acceptance cuts of $E_\gamma>10$ GeV and
      $|\cos\theta_\gamma|<0.9$, correspondingly. Fig.~\ref{photon} 
      agrees very well with the analogous Fig.~3 of \cite{Cheung:1999zw}
      \footnote{Note that the calculation in \cite{Cheung:1999zw}
        is based on Feynman rules from an earlier
        version of \cite{Han:1998sg}, leading to an intermediate
        result that is too small by a factor 
        of $1/2$~\cite{Mirabelli:1998rt,Tikhonin:2002ie}.
        However, along with the correction of that factor of $1/2$, the 
        revised version of \cite{Han:1998sg} also redefined
        the scale $M_S$ in such a way that the physical cross sections 
        for spin-2 KK gravitons remained unchanged 
        (after integration over the KK tower).
        Thus, Fig.~3 of \cite{Cheung:1999zw} can be safely
        used for comparison purposes here.}.
\item Finally, we calculated the decay widths of massive 
      KK particles. For spin-2 KK states we agree with the results
      of \cite{Han:1998sg}. For spin-0 particles 
      we agree with the results in the original printed
      version of \cite{Han:1998sg}, while in its revised
      version the widths are too
      large by a factor of 2. 
\end{itemize}
This implementation of the ADD model
will be available in the upcoming version 2.0 of {\tt AMEGIC++}.

\section{Conclusions}
\label{sec:conclusions}

The idea that quantum gravity can be relevant already 
at the TeV scale is extremely attractive -- 
to theorists and experimentalists alike.
Since gravity is universal,
there exist many potential graviton signatures, 
but so far only the simplest ones have been explored. 
The calculations become prohibitively complicated 
and the need for automation soon becomes evident.
Thus the main purpose of this paper was to develop 
the helicity formalism for spin-2 particles, which
could then be used in an automatic event generator,
capable of calculating 
arbitrary processes with gravitons.  

Having completed this first task, we proceeded to implement
the ADD model of large extra dimensions
in {\tt AMEGIC++}. 
The large extra dimensions paradigm will soon be tested 
in collider experiments at the Tevatron, LHC and NLC. 
We believe that having a simulation tool like
{\tt AMEGIC++} at our disposal will greatly facilitate the 
study of new physics related to quantum gravity 
or extra dimensions. With the general formalism at hand,
it is now possible to generalize our implementation 
to models with warped extra dimensions as well.

\bigskip

\acknowledgments
We thank T.~Han, J.~Lykken, M.~Perelstein, D. Rainwater 
and F.~Tikhonin for comments.
TG wants to thank the DAAD for funding and the Physics Department of the
University of Florida for warm hospitality and additional support.  
FK gratefully acknowledges funding through the EC 5th Framework Programme
under contract number HPMF-CT-2002-01663. The work of KM is supported by 
the US Department of Energy under grant DE-FG02-97ER41209. AS thanks 
DESY/BMBF for financial support. SS wants to express his gratitude
to the GSI Darmstadt for funding. 
\appendix
\section{Appendix: \ Vertex Functions}
\label{Vertex}
\renewcommand{\theequation}{A.\arabic{equation}}
\setcounter{equation}{0}

The following shorthands are used:
\bea
Y_{0} &=&Y(k_0,\lambda_0,k_0',\lambda_0',c_L,c_R)\,,\nonumber \\
X_{01}&=&X(k_0,\lambda_0,k_1,k_0',\lambda_0',c_L,c_R)\,,\nonumber \\
X_{0\kappa}&=&X(k_0,\lambda_0,\epsilon^{\kappa},k_0',\lambda_0',c_L,c_R)
\,,\nonumber \\
Z_{01}&=&Z(k_0,\lambda_0,k_0',\lambda_0',
k_1,\lambda_1,k_1',\lambda_1',c_L^0,c_R^0,c_L^1,c_R^1)\,.\nonumber
\eea
All spin-2 propagators are ``cut'' (see \ref{SecS1Prop}, \ref{SecS2Prop}), 
while the spin-1 propagators
are attached to a fermion line. For the case when they are attached to
something else in a diagram (and for massive vector bosons), 
they must be ``cut'' as well. This is done by simple replacements 
(e.g. to cut the boson line with momentum $p_0$):
\bea
X_{01} &\rightarrow& \epsilon_{\mu}^{\lambda}k_1^{\mu}\,,\nonumber \\
Z_{01} &\rightarrow& X_{1\lambda}\,,\nonumber 
\eea
where $\lambda$ labels the polarization modes of the cut line.
%
%
\begin{figure}[h!]
\begin{center}
{
\unitlength=1.0 pt
\SetScale{1.0}
\SetWidth{0.7}      
\normalsize    
{} \qquad\allowbreak
\begin{picture}(300,100)(0,0)
\SetColor{Blue}
\ArrowLine(50,50)(0,100)
\ArrowLine(0,0)(50,50)
\SetColor{Red}
\Photon(50,50)(100,50){3}{5}
\Photon(50,50)(100,50){-3}{5}
\Text(10,30)[c]{\Blue{$\Psi(k_{0})$}}
\Text(10,70)[c]{\Blue{$\Psi(k_{0}')$}}
\Text(70,60)[c]{\Red{$\epsilon^{\lambda}_{\mu},\epsilon^{\kappa}_{\nu}$}}
\Text(110,70)[l]{\Black{Spin-2 KK: $\; -\frac{i}{8}\kappa\;\mathit{FFT}$}}
\Text(110,30)[l]{\Black{Spin-0 KK: $\; i\frac{3}{4}\omega\kappa\;\mathit{FFGS}$}}
\end{picture}
}
\caption{\label{fig1} The fermion--fermion--KK vertex.}
\end{center}
\end{figure}
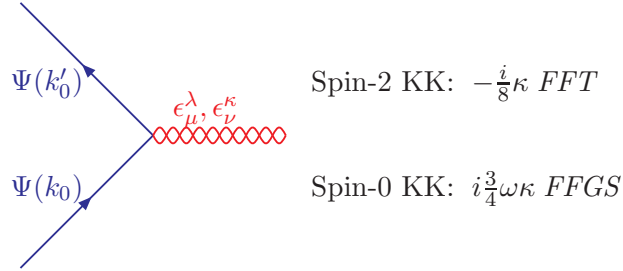

\begin{eqnarray}
FFT &=& X_{0\lambda}\left(k_{0}\!\cdot\!\epsilon^{\kappa}-
k_{0}'\!\cdot\!\epsilon^{\kappa}\right)
+X_{0\kappa}\left(k_{0}\!\cdot\!\epsilon^{\lambda}-
k_{0}'\!\cdot\!\epsilon^{\lambda}\right)\,,\nonumber \\
& & -2\,\epsilon^{\lambda}\!\cdot\!\epsilon^{\kappa}\left(
X\left(k_{0},k_{0},k_{0}'\right)-X\left(k_{0},k_{0}',k_{0}'\right)-
2 m_{\Psi} Y_{0}\right)\\
FFGS &=& X\left(k_{0},k_{0},k_{0}'\right)-X\left(k_{0},k_{0}',k_{0}'\right)-
\frac{8}{3} m_{\Psi} Y_{0}\,.
\end{eqnarray}

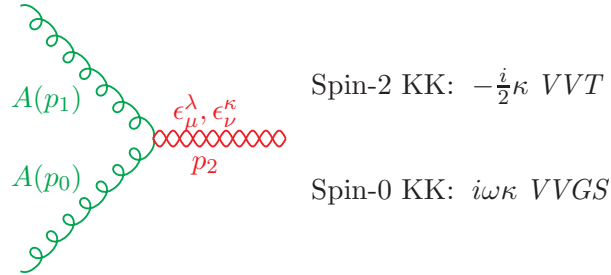
\begin{figure}[h!]
\begin{center}
{
\unitlength=1.0 pt
\SetScale{1.0}
\SetWidth{0.7}      
\normalsize    
{} \qquad\allowbreak
\begin{picture}(300,100)(0,0)
\SetColor{Green}
\Gluon(0,100)(50,50){3}{7}
\Gluon(50,50)(0,0){3}{7}
\SetColor{Red}
\Photon(50,50)(100,50){3}{5}
\Photon(50,50)(100,50){-3}{5}
\Text(10,35)[c]{\Green{$A(p_{0})$}}
\Text(10,65)[c]{\Green{$A(p_{1})$}}
\Text(70,60)[c]{\Red{$\epsilon^{\lambda}_{\mu},\epsilon^{\kappa}_{\nu}$}}
\Text(70,40)[c]{\Red{$p_{2}$}}
\Text(110,70)[l]{\Black{Spin-2 KK: $\; -\frac{i}{2}\kappa\;\mathit{VVT}$}}
\Text(110,30)[l]{\Black{Spin-0 KK: $\; i\omega\kappa\;\mathit{VVGS}$}}
\end{picture}
}
\caption{\label{fig2} The two-gauge-boson--KK vertex.}
\end{center}
\end{figure}

\begin{eqnarray}
VVT &=& \left(m_{A}^{2}+p_{0}\!\cdot\!p_{1}\right)\left(X_{0\lambda}X_{1\kappa}
+X_{1\lambda}X_{0\kappa}-\epsilon^{\lambda}\!\cdot\!\epsilon^{\kappa}\,Z_{01}
\right)\nonumber \\
& & +\left(X_{01}+X_{00}\right)\left(X_{10}+X_{11}\right)
\epsilon^{\lambda}\!\cdot\!\epsilon^{\kappa}
+Z_{01}\left(p_{0}\!\cdot\!\epsilon^{\kappa}\;p_{1}\!\cdot\!\epsilon^{\lambda}
+p_{1}\!\cdot\!\epsilon^{\kappa}\;p_{0}\!\cdot\!\epsilon^{\lambda}\right)\nonumber \\
& & -\left(X_{01}+X_{00}\right)\left(X_{1\kappa}\,p_{0}\!\cdot\!\epsilon^{\lambda}+
X_{1\lambda}\,p_{0}\!\cdot\!\epsilon^{\kappa}\right)\nonumber \\
& & -\left(X_{10}+X_{11}\right)\left(X_{0\kappa}\,p_{1}\!\cdot\!\epsilon^{\lambda}+
X_{0\lambda}\,p_{1}\!\cdot\!\epsilon^{\kappa}\right)\,,\\
VVGS &=& m_{A}^{2}Z_{01}+X_{00}X_{12}+X_{02}X_{11}\,.
\end{eqnarray}


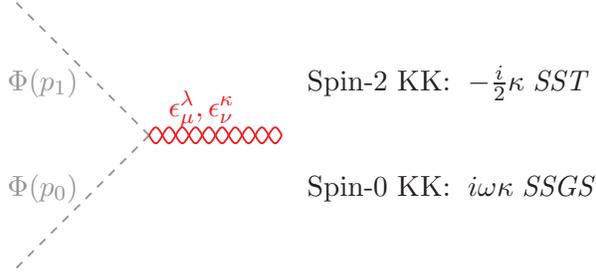
\begin{figure}[h!]
\begin{center}
{
\unitlength=1.0 pt
\SetScale{1.0}
\SetWidth{0.7}      
\normalsize    
{} \qquad\allowbreak
\begin{picture}(300,100)(0,0)
\SetColor{Gray}
\DashLine(0,100)(50,50){3}
\DashLine(50,50)(0,0){3}
\SetColor{Red}
\Photon(50,50)(100,50){3}{5}
\Photon(50,50)(100,50){-3}{5}
\Text(10,30)[c]{\Gray{$\Phi(p_{0})$}}
\Text(10,70)[c]{\Gray{$\Phi(p_{1})$}}
\Text(70,60)[c]{\Red{$\epsilon^{\lambda}_{\mu},\epsilon^{\kappa}_{\nu}$}}
\Text(110,70)[l]{\Black{Spin-2 KK: $\; -\frac{i}{2}\kappa\;\mathit{SST}$}}
\Text(110,30)[l]{\Black{Spin-0 KK: $\; i\omega\kappa\;\mathit{SSGS}$}}
\end{picture}
}
\caption{\label{fig3} The scalar--scalar--KK vertex.}
\end{center}
\end{figure}

\begin{eqnarray}
SST &=& \left(m_{\Phi}^{2}+p_{0}\!\cdot\!p_{1}\right)
\epsilon^{\lambda}\!\cdot\!\epsilon^{\kappa}-
p_{0}\!\cdot\!\epsilon^{\lambda}\;p_{1}\!\cdot\!\epsilon^{\kappa}-
p_{1}\!\cdot\!\epsilon^{\lambda}\;p_{0}\!\cdot\!\epsilon^{\kappa}\,,\\
SSGS &=& -2 m_{\Phi}^{2}+p_{0}\!\cdot\!p_{1}\,.
\end{eqnarray}

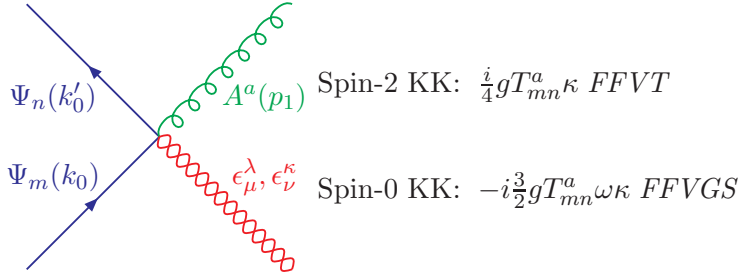
\begin{figure}[h!]
\begin{center}
{
\unitlength=1.0 pt
\SetScale{1.0}
\SetWidth{0.7}      
\normalsize    
{} \qquad\allowbreak
\begin{picture}(300,100)(0,0)
\SetColor{Blue}
\ArrowLine(50,50)(0,100)
\ArrowLine(0,0)(50,50)
\SetColor{Green}
\Gluon(50,50)(100,100){3}{7}
\SetColor{Red}
\Photon(50,50)(100,0){3}{7}
\Photon(50,50)(100,0){-3}{7}
\Text(10,35)[c]{\Blue{$\Psi_m(k_{0})$}}
\Text(10,65)[c]{\Blue{$\Psi_n(k_{0}')$}}
\Text(90,65)[c]{\Green{$A^a(p_{1})$}}
\Text(90,35)[c]{\Red{$\epsilon^{\lambda}_{\mu},\epsilon^{\kappa}_{\nu}$}}
\Text(110,70)[l]{\Black{Spin-2 KK: 
$\; \frac{i}{4}gT_{mn}^{a}\kappa\;\mathit{FFVT}$}}
\Text(110,30)[l]{\Black{Spin-0 KK: 
$\; -i\frac{3}{2}gT_{mn}^{a}\omega\kappa\;\mathit{FFVGS}$}}
\end{picture}
{} \qquad\allowbreak
}
\caption{\label{fig4} The fermion--fermion--gauge--boson--KK vertex.}
\end{center}
\end{figure}

\begin{eqnarray}
FFVT &=& X_{0\kappa}X_{1\lambda}+X_{1\kappa}X_{0\lambda}-
2\,\epsilon^{\lambda}\!\cdot\!\epsilon^{\kappa}Z_{01}\,, \\
FFVGS &=& Z_{01}\,.
\end{eqnarray}

\newpage

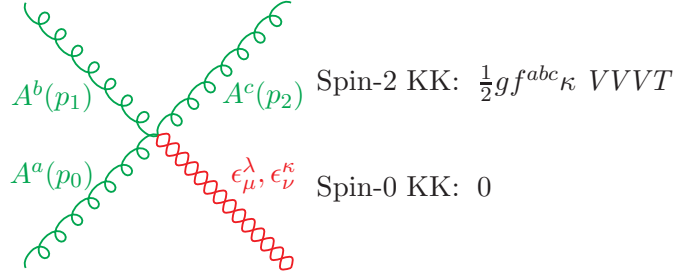
\begin{figure}[h!]
\begin{center}
{
\unitlength=1.0 pt
\SetScale{1.0}
\SetWidth{0.7}      
\normalsize    
{} \qquad\allowbreak
\begin{picture}(300,100)(0,0)
\SetColor{Green}
\Gluon(50,50)(0,100){3}{7}
\Gluon(0,0)(50,50){3}{7}
\Gluon(50,50)(100,100){3}{7}
\SetColor{Red}
\Photon(50,50)(100,0){3}{7}
\Photon(50,50)(100,0){-3}{7}
\Text(10,35)[c]{\Green{$A^a(p_{0})$}}
\Text(10,65)[c]{\Green{$A^b(p_{1})$}}
\Text(90,65)[c]{\Green{$A^c(p_{2})$}}
\Text(90,35)[c]{\Red{$\epsilon^{\lambda}_{\mu},\epsilon^{\kappa}_{\nu}$}}
\Text(110,70)[l]{\Black{Spin-2 KK: $\; \frac{1}{2}gf^{abc}\kappa\;
\mathit{VVVT}$}}
\Text(110,30)[l]{\Black{Spin-0 KK: $\; 0$}}
\end{picture}
{} \qquad\allowbreak
}
\caption{\label{fig5} The triple--gauge--boson--KK vertex.}
\end{center}
\end{figure}
 
\begin{eqnarray}
VVVT &=& A_{012}+A_{120}+A_{210}\,, \\ \nonumber \\
A_{012} &=& \left(X_{0\lambda}X_{1\kappa}+X_{1\lambda}X_{0\kappa}-
\epsilon^{\lambda}\!\cdot\!\epsilon^{\kappa}\,Z_{01}\right)
\left(X_{20}-X_{21}\right)\nonumber \\
& & + Z_{01}\left[X_{2\lambda}
(p_{0}\!\cdot\!\epsilon^{\kappa}-p_{1}\!\cdot\!\epsilon^{\kappa})+
X_{2\kappa}(p_{0}\!\cdot\!\epsilon^{\lambda}-p_{1}\!\cdot\!\epsilon^{\lambda})\right]
\,.\nonumber
\end{eqnarray}

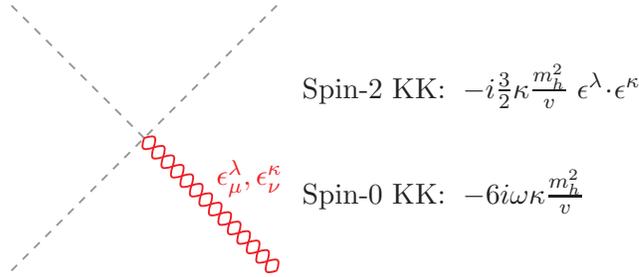
\begin{figure}[h!]
\begin{center}
{
\unitlength=1.0 pt
\SetScale{1.0}
\SetWidth{0.7}      
\normalsize    
{} \qquad\allowbreak
\begin{picture}(300,100)(0,0)
\SetColor{Gray}
\DashLine(0,100)(50,50){3}
\DashLine(100,100)(0,0){3}
\SetColor{Red}
\Photon(50,50)(100,0){3}{7}
\Photon(50,50)(100,0){-3}{7}
\Text(90,35)[c]{\Red{$\epsilon^{\lambda}_{\mu},\epsilon^{\kappa}_{\nu}$}}
\Text(110,70)[l]{\Black{Spin-2 KK: $\; -i\frac{3}{2}\kappa\frac{m_{h}^{2}}{v}\;
\epsilon^{\lambda}\!\cdot\!\epsilon^{\kappa}$}}
\Text(110,30)[l]{\Black{Spin-0 KK: $\; -6i\omega\kappa\frac{m_{h}^{2}}{v}$}}
\end{picture}
}
\caption{\label{fig6} The triple--scalar--KK vertex.}
\end{center}
\end{figure}

\listoftables           
\listoffigures          


\end{document}